\documentclass[12pt,preprint]{aastex}
\usepackage{amsmath, amsthm}
\shorttitle{Relativistic Boltzmann}
\shortauthors{Wolfe}

\begin{document}

\title{Covariant Kinetic Theory with an Application to the Coma Cluster}

\author{Brandon Wolfe$^{1}$ and Fulvio Melia$^{1,2}$}
\affil{$^1$ Physics Department, The University of Arizona, Tucson, AZ 85721}
\affil{$^2$ Steward Observatory, The University of Arizona, Tucson, AZ 85721}

\begin{abstract}
In this paper, we introduce a novel solution to the covariant Landau equation for a pure
electron plasma.  The method conserves energy and particle number, and reduces smoothly 
to the Rosenbluth potentials of non-relativistic theory. In addition, we find that a fully 
relativistic plasma equilibrates in only 1/100th of a Spitzer time---much faster than
in the non-relativistic limit---a factor of significant import to situations in which 
distortions to a Maxwellian distribution are produced by anomalous methods of acceleration.  
To demonstrate the power of our solution in dealing with hot, astrophysical plasmas, we use 
this technique to show that one of the currently considered models---continuous stochastic 
acceleration---for the hard X-ray emission in the Coma cluster actually cannot work because 
the energy gained by the particles is distributed to the {\it whole} plasma on a time scale 
much shorter than that of the acceleration process itself. 

\end{abstract}

\keywords{acceleration of particles --- galaxies: clusters: individual (Coma) ---
plasmas --- radiation mechanisms: non-thermal --- relativity --- X-rays: galaxies}

\section{Introduction}
The need for a self-consistent, practical transport theory that can handle dynamic
populations of relativistic particles extends across many disciplines in physics. 

Most obviously, astrophysical plasmas in the presence of strong gravitational and/or magnetic fields are often out 
of thermal equilibrium when the relevant dynamic time scale (e.g., associated with the process of 
accretion onto a compact object) is short compared to the time required for the particles to interact
internally with each other.  Situations in which this may occur include solar flares (Petrosian and
Liu 2003), accretion onto supermassive black holes in the nuclei of active galaxies (see, e.g., Melia and
Falcke 2001), and re-acceleration of already relativistic particles in shocks produced during the 
merger of galaxies (Brunetti et al.
2004).  In all cases, the resultant photon spectrum produced by the energized particles is most simply
explained in terms of a non-Maxwellian distribution.

Less directly, an argument can be made that the two dominant particle species (say, electrons and
protons in a fully ionized Hydrogen plasma) separate in energy space, leading to a situation in which
the ion temperature is larger than that of the electrons (see, e.g., Shapiro, Lightman, and
Eardley 1976). 
However, whether such a two-temperature plasma may be maintained against 
collisional re-equilibration is still actively debated, and may be critical to the question of
how a black hole accretes from its environment.  For example, there is now very good evidence that
the gas orbiting the supermassive black hole, Sgr A*, at the center of our Galaxy attains a temperature 
in excess of $10^{10}$ K (Liu and Melia 2001), but it is still not understood whether this optically 
thin plasma maintains a single temperature. In addition, nonthermal emission in Sgr A* (see, e.g., Ozel, 
Psaltis, and Narayan 2000;  Yuan, Quataert, and Narayan 2003) has been invoked to explain
its low-frequency radio spectrum in accretion-dominated models.

These clear astrophysical applications should be aided by recent experiments with 
laser-heated plasmas, which are now probing the energy regimes of interest. 
Phenomena unique to relativistic distributions, 
such as their tendency to limit heat flux (Bell et al. 1981; Rickard et al. 1989), or the ability of
rapid acceleration to create bi-Maxwellian electron distributions (Guethlein et al. 1996)
have now been observed in laboratory plasmas. Experimental constraints arising from both the
astrophysical and laboratory contexts, acting in concert, should yield a solid, well-tested theory.

Astrophysicists have had a theory for relativistic energy transport for some time. 
Dewar et al. (1979) were followed by Dermer 
and Liang (1989) in calculating the energy exchanged between particles during 
a relativistic binary collision.  Nayakshin and Melia (1998) used these 
coefficients as the basis of a Fokker-Planck equation in energy. But, as 
Dewar et al. note, ``virtually all of the quantities that were invariant in 
the non-relativistic treatment cease to be so in the relativistic formulation. 
Much care (must) be taken to show how the various physical quantities...
transform." The resulting theory is not as transparent as one would like.

In contrast, the strategy we adopt here follows Beliaev and Budker (1955) in 
formulating kinetic theory in a covariant form valid in any frame of reference. 
This approach is simpler, since the frame invariants are obvious.
It is also general: we present a solution that handles anisotropic
or radically out-of-equilibrium distributions, and can be used to 
find electric and thermal conduction coefficients as well as energy 
exchange. This approach also conveniently and elegantly reduces to the 
well-known Rosenbluth (1957) theory in the non-relativistic limit.

Relativistic distributions obey \emph{qualitatively} different physics, since the 
covariant phase volume must incorporate energy to form an eight-dimensional phase space.
But the covariant theory is also \emph{quantitatively} sufficiently different to 
shape observations.  We show here that a fully relativistic distribution equilibrates 
in a tenth to a hundredth the time given by non-relativistic theory.
The heat flux in a relativistic 
distribution, relative to that for the non-relativistic case, is inhibited by a factor 
of $10^{-2}-10^{-1}$. This is due to a drift velocity which approaches the speed of light
asymptotically.  Thus, situations where a nonthermal tail is likely to be 
present (see, e.g., Blasi 2000; we will examine this situation in greater detail below) 
require a covariant formulation.

Solutions to covariant kinetic theory have existed for some time, too. 
Braams and Karney (1987) made a spherical expansion of 
the kinetic kernel and
found relativistic conductivities (1989), 
but only in very limited circumstances. This was followed by attempts at solution via a Chapman-Enskog
expansion (Mohanty and Baral 1996), Legendre expansion (Honda 2003), Grad expansion (Muronga 2004),
and numerical integration (Shoucri and Shakarofsky 1994).

But, to our knowledge, none of these solutions is six dimensional---the ability to solve three 
dimensional integrals like (Eq. 37 below) has only existed for a short time (Hahn 2004). 
Manheimer et al. (1997) have called the 
self-consistent evaluation of even the simpler non-relativistic coefficients ``unfeasible". 

Many approach the Fokker-Planck equation with implicit differencing, which is relatively 
difficult to modify with additional forces (e.g., in a particle-in-cell simulation).
Instead, we follow the original suggestion of Jones et al. (1996) and the subsequent application of 
Habib et al. (2000) to model collisions as Brownian motion. A single particle's motion is observed at 
discrete times, and the affect of multiple collisions accumulated over this time is 
represented as dynamical friction and noise. The distribution is approximated by some 
small fraction of the actual number of particles, all moving according to self-consistent
kinetic forces. The simulation thus solves single-body statistical dynamics, many times.  

The desire for a covariant theory that is simple to apply and reduces naturally to the non-relativistic
version is far from academic. As a concrete---and important---example, we show that our theory eliminates 
the possibility that stochastic acceleration is responsible for the hard X-ray emission in the Coma cluster.
Before that, however, we present the relativistic kinetic theory in \S\ 2, and then describe the
method of solution in covariant form in \S\ 3.  The application will be made in \S\ 4.

\section{Relativistic Kinetic Theory: Background}

We begin by discussing the development of the relativistic kinetic equation somewhat
pedagogically, given the fact that its usage in the analysis of high-energy plasmas
in astrophysics has been infrequent at best.  Landau (1936) first approached the problem
by showing that the Boltzmann collision integral can be written 
as a flux in velocity space,
\begin{equation}
 C = - \frac{\partial}{\partial v_\alpha}S_\alpha\;,
\end{equation}
thereby writing the Boltzmann equation as a conservation equation for the 
single particle phase space distribution,
\begin{equation}
 \frac{\partial}{\partial t} f + \{ H,f \} = \bigtriangledown  \centerdot \mathbf{S}\;,
\end{equation}
where $f$ is the single particle phase space \emph{number} density.  The Poisson bracket $\{H,f\}$ represents
a small change in the shape of a volume of phase space (which nevertheless conserves number), 
the term $\bigtriangledown \centerdot \mathbf{S}$ a flux of particles 
out of it. But what is $S$? To begin with, 
note that the number of collisions occurring in unit time between a particle with velocity
$\mathbf{v}$ and particles with velocity $\mathbf{v^\prime}$ in the range $d^3 v^\prime $ is
\begin{equation}
 w f( \mathbf{v} ) f( \mathbf{v^\prime} )\; d^3q\;d^3v^\prime\;,
\end{equation}
where $q$ is the amount of velocity exchanged.  Here, $w$, the probability that 
$\mathbf{v}$ will be taken into $\mathbf{v+q}$, can be expressed evenly as 
$w( \mathbf{v} + \mathbf{q}/2, \mathbf{v^ \prime} - \mathbf{q}/2; \mathbf{q})$. 
This ensures that $w$ is symmetrical with respect to $\mathbf{q}$---i.e., that detailed balance is maintained.

The particle flux $F=F^+-F^-$ is found by subtracting the number of particles entering a volume of velocity space
from the left,
\begin{equation}
F^- \equiv\int d^3 q \int d^3 v^\prime \int ^{p_\alpha} _{p_\alpha - q_\alpha}
	w f( \mathbf{v} ) f^\prime ( \mathbf{v^\prime} )\; dv_\alpha
\end{equation}
from that exiting to the right,
\begin{equation}
F^+ \equiv\int d^3 q \int d^3 v^\prime \int ^{p_\alpha} _{p_\alpha - q_\alpha}
	w f( \mathbf{v+q} ) f^\prime ( \mathbf{v^\prime-q} )\; dv_\alpha\;.
\end{equation}
This leads to an argument 
$f( \mathbf{v} ) f^\prime ( \mathbf{v^\prime} ) - f( \mathbf{v+q} ) f^\prime ( \mathbf{v^\prime-q} )$
which can be expanded about small $q$, giving the required $S$, the so-called Landau collision integral,
\begin{equation}
 C = \frac{\partial}{\partial v_\alpha}S_\alpha = 
\frac{\partial}{\partial v_\alpha}\sum_\beta\int\bigg[f(\mathbf{v})\frac{\partial f^\prime(\mathbf{v}^\prime)}
{\partial v^\prime_\beta}-f^\prime(\mathbf{v}^\prime)\frac{\partial f(\mathbf{v})}{\partial v_\beta}\bigg]
		B_{\alpha\beta}\; d^3v\;,
\end{equation}
where $w$ is expressed in terms of the collision cross section,
\begin{equation}
 w\; d^3 q = | \mathbf{v} - \mathbf{v^\prime} |\; d \sigma\;,
\end{equation}
and where
\begin{equation}
 B_{\alpha \beta} \equiv \frac{1}{2} \int q_\alpha q_\beta | \mathbf{v}-\mathbf{v^\prime}|\; d \sigma\;.
\end{equation}
Finally, since momentum conservation dictates that the change $\mathbf{q}$ is
perpendicular to the relative velocity $|\mathbf{v}-\mathbf{v^\prime}|$, it must be that
\begin{equation}
 B_{\alpha \beta}\,(v_\beta - v_\beta^\prime) = 0\;.
\end{equation}
And since $ B_{\alpha \beta} $ can only depend on the vector $|\mathbf{v}-\mathbf{v^\prime}|$,
the tensor must take the form
\begin{equation}
 B_{\alpha \beta} = \frac{1}{2} B \bigg[ \delta_{\alpha \beta} - 
		\frac {(v_\alpha - v_\alpha^\prime)(v_\beta- v_\beta^\prime)}
		{(\mathbf{v}-\mathbf{v^\prime})^2 } \bigg]\;,
\end{equation}
where
\begin{equation}
 B \equiv B_{\alpha \alpha} = \frac{1}{2}\int q^2 | \mathbf{v}-\mathbf{v^\prime}|\; d \sigma\;.
\end{equation}
Thus, integrating Equation (11) with a Rutherford cross-section for $ d \sigma$, one gets
\begin{equation}
 B_{\alpha\beta} = \frac{e^2\,{e^{\prime}}^2}{8 \pi \epsilon_0^2 m^2}\,\ln\Lambda\;
			\mathbf{U}(\mathbf{v},\mathbf{v^\prime})\;,
\end{equation}
where $\mathbf{U}(\mathbf{v},\mathbf{v^\prime})$ is the collision kernel,
\begin{equation}
 \mathbf{U}(\mathbf{v},\mathbf{v^\prime}) \equiv \frac{|\mathbf{v}-\mathbf{v^\prime}|^2 \delta_{\alpha\beta} 
		- (v_\alpha - v_\alpha^\prime)
	(v_\beta-v_\beta^\prime)}{|\mathbf{v}-\mathbf{v^\prime}|^3}\;,
\end{equation}
\\
and $\alpha, \beta$ are Cartesian indices. In what follows, we define
$A\equiv n {e^2 {e^{\prime}}^2}/{8 \pi \epsilon_0^2 m^2}\,\ln\Lambda$ and $l\equiv \sqrt{{m}/{2kT}}$
for succinctness. In c.g.s. units, 
$A \equiv 8 \pi e^2 {e^{\prime}}^2 n \text{ln} \Lambda / m^2$.

With one integration by parts, the collision integral can be written 
\\
\begin{equation}
C = \frac{\partial}{\partial v_\alpha}\bigg[D_{\alpha\beta}\frac{\partial 
f(\mathbf{v})}{\partial v_\beta}-F_\alpha\, f(\mathbf{v})\bigg]\;,
\end{equation}
\\
where
\begin{equation}
 D_{\alpha\beta} \equiv A
	\int\mathbf{U}(\mathbf{v},\mathbf{v^\prime})f(\mathbf{v^\prime})\; d^3 v^\prime\;,
\end{equation}
and
\begin{equation}
 F_\alpha \equiv A
	\sum_\beta\int\bigg[ \frac{\partial}{\partial v^\prime_\beta}\mathbf{U}(\mathbf{v},
\mathbf{v^\prime}) \bigg] f(\mathbf{v^\prime})\; d^3 v^\prime\;,
\end{equation}
\\
which is the Fokker-Planck equation seen commonly in the west. Equations
(15) and (16) can be expressed in a simpler form by observing that the kernel $\mathbf{U}(\mathbf{v},\mathbf{v^\prime})$
obeys the relations
\\
\begin{equation}
  U_\alpha = \frac{\partial^2 |\mathbf{v}-\mathbf{v^\prime}|}{\partial v_\alpha \partial v_\beta}\;,
\end{equation}
and
\begin{equation}
  \frac{\partial}{\partial v_\alpha^\prime}\, U_\alpha = \frac{-2 \partial |\mathbf{v}-
\mathbf{v^\prime}|^{-1} }{\partial v_\alpha}\;,
\end{equation}
\\
thus allowing the coefficients to be expressed as
\\
\begin{equation}
 D_{\alpha\beta} = -A
		\frac{\partial^2}{\partial v_\alpha \partial v_\beta}\bigg[
		\int f |\mathbf{v}-\mathbf{v^\prime}| d^3 v^\prime\bigg]\;,
\end{equation}
and
\begin{equation}
 F_\alpha = -A
		\frac{\partial}{\partial v_\alpha}\bigg[2 \int f |\mathbf{v}-
\mathbf{v^\prime}|^{-1}\;d^3 v^\prime\bigg]\;.
\end{equation}

The advantage of this phrasing (Rosenbluth et al. 1957) is that the terms within the brackets contain
elements in common with the Poisson equation, for which sophisticated methods of solution exist
(e.g., expansion by spherical harmonics and FFT convolution). When $f(\mathbf{v})$ is a 
Maxwellian, a harmonic expansion gives

\begin{equation}
 F = -A l^2 \,\left(1+\frac{m}{m^\prime}\right)\, G(l v) \;,
\end{equation}

\begin{equation}
 D_\bot = \frac{A}{v}\, \bigg[ \Phi(lv) - G(lv) \bigg]\;,
\end{equation}
and
\begin{equation}
 D_{||} = \frac{A}{v}\, G(lv)\;,
\end{equation}
where $D_{||}$ $(D_\bot)$ is the component of the diffusion tensor parallel (perpendicular) to the
 particle's velocity in the co-moving frame, $\Phi$ is the error function, and 
$G(x) \equiv \Phi(x)-x\Phi^\prime(x) / 2x^2$. Since $dv/dv_i = v_i/v$, the diffusion 
tensor fills out as
\begin{equation}
D_{\alpha \beta} = \frac{v_\alpha v_\beta}{v^2}\, \bigg[ D_{||}-\frac{1}{2} D_\bot \bigg] + \frac{1}{2} 
\delta_{\alpha \beta}\, D_\bot\;.
\end{equation}
Equations (21) thru (24) present a fairly complete non-relativistic kinetic theory.

The time $t_d$ required for a particle to be deflected by 90 degrees is a good measure of how 
long it will take a non-thermal injection to equilibrate (Spitzer 1962). This time is given by the condition 
\begin{equation}
D_\bot\, t_s = v^2\;.
\end{equation}
Thus, for particles whose root mean square velocity is that of the group,
$v = \sqrt{ {3kT}/{m} }$, we have $lv = 1.225$, $D_\bot = 0.714 A\, $ (m/s)$^2$/s, and
\begin{equation}
 t_s =  \bigg( \frac {3 kT}{m} \bigg)^{3/2} {1\over (0.714 A)}\;\, \hbox{seconds}\;.
\end{equation}
A slowing-down viscous time scale can be similarly defined. 
One of our primary goals is to find a relativistically correct version of Equation (26)---
the majority of problems (even apparently difficult ones, such as nonthermal radiation
from galaxy clusters) can be resolved with a glance at this simple formula.

Chandrasekhar (1957) approached the problem quite differently. 
Consider replacing the truly discrete many-body
potential $ \Phi = \sum_i e_i e^\prime_i / |\mathbf{r}-\mathbf{r^\prime}|$ with the smoothed version
$ \Phi (\mathbf{r}) = \int e \rho ( \mathbf{r^\prime}) /  |\mathbf{r}-\mathbf{r^\prime}| d^3 r^\prime $.
 A distribution evolving only due to such a smooth, mean field force is `collisionless'---i.e., no particle 
ever leaves its volume of (single particle)  phase space.  But in moving to an integrable charge
distribution, we have lost track 
of small fluctuations in the number of particles in a given neighborhood.
  
Accumulated over a small time, these fluctuations cause a random kink in a particle's path. Alone, this
kink diffuses the probability of finding the particle with a particular velocity until all velocities are 
equally likely.  Since,  however, a Maxwellian distribution for this probability inevitably sets in, 
Chandrasekhar reasoned that an associated viscous term must exist to bring diffusing particles back to the 
group, so that individual particles move through phase space according to the equations of motion
\\
\begin{equation}
 {d\mathbf{x}\over dt} = \mathbf{v}
\end{equation}
\begin{equation}
 {d\mathbf{v}\over dt} = \frac{\mathbf{F}}{m}+\mathbf{F_d}+\mathbf{Q}\centerdot\mathbf{\Gamma}(t)\;,
\end{equation}
\\
where $\mathbf{F}$ is a smooth mean field term (plus any external forces),
$\mathbf{F_d}$ is dynamical friction that occurs when more
particles hit the front end of a moving particle than the back, 
$\mathbf{Q}$ is the `square root' of the diffusion tensor,
 $D_{i j} = Q_{i k} Q_{j k}$,
and $\mathbf{\Gamma}(t)$ is a trivariate Gaussian random vector with
\begin{equation}
 \langle \Gamma_i (t) \rangle = 0\;,
\end{equation}
\begin{equation}
 \langle \Gamma_i (t) \Gamma_j (t) \rangle = \delta_{i j}\, \delta (t-t^\prime)\;.
\end{equation}
We show in Appendix A why Equations (2) and (28) are equivalent.

Equations (27) and (28) reappropriate kinetic theory as a dynamics that is conceptually 
different than Hamilton's: Analytical dynamics follows the continuous motions of each of $N$ 
particles in a 2N-dimensional phase space; it conserves energy, maintains the number of particles 
in a volume of phase space as a constant, and is reversible.  Analytical dynamics has been described 
as ``a gradual unfolding of a contact transformation." 

By contrast, statistical dynamics follows the motion of particles accumulated over some coarse time 
$\triangle t$---indeed, since $\sqrt{ \bar{ | \triangle \mathbf{v} |^2 } } / \triangle t \rightarrow \infty$
as $\triangle t \rightarrow 0$,
statistical dynamics is not even defined continuously. Statistical dynamics ignores all but the 6 phase space
coordinates of the particle in question, representing the effect of the other degrees of freedom 
as friction and noise---these terms cause a flux of particles through a surface of the \emph{single particle} phase space.
Statistical dynamics does not conserve energy, having at its heart the dissipative force $F_d$, yet this force
is precisely what is required to bring the assembly as a whole to a Maxwellian distribution characterized
by a constant energy. Statistical dynamics is irreversible. Chandrasekhar described it as the ``gradual
unfolding of a Markoff chain."
 
Solving Equation (1) by moving many particles according to the equations of motion (27, 28) is
sometimes called a `Monte Carlo' approach to kinetic theory.  It is important to stress the 
historical development here to avoid the tincture of a computational trick that this name implies: 
the Langevin Equation (28) is equivalent to the Fokker-Planck equation, and implies no further approximation 
provided collisions are short (see Eq. 30).  
It can be developed on physical grounds entirely separate from the 
Boltzmann equation, and in fact was developed some 14 years before a successful solution of the 
Fokker-Planck equation appeared.

While we will use Equations (27) and (28) to solve the evolution of the distribution, we return to Equation (2) 
to derive the relativistically correct kinetic coefficients.  The crucial distinction between non-relativistic 
theory and the relativistic version that follows is that the function
\begin{equation}
f(\mathbf{p}) = \int f(t, \mathbf{r}, \mathbf{p} )\; d^3 x
\end{equation}
is \emph{not} a relativistically invariant 4-scalar, since in different coordinate systems 
different particles will be considered
to be located simultaneously in a given volume. The integrals
$N = \int f(t,\mathbf{r}, \mathbf{p})\; d^3 p$, and $\mathbf{i} = \int \mathbf{v} f(t,\mathbf{r}, \mathbf{p})\; d^3 p$,
however, \emph{do} form a 4-vector: 
\begin{equation}
i^k = (cN, \mathbf{i}) = c^2 \int (p^k f/\epsilon)\; d^3 p\;.
\end{equation}
Here $ d^3 p/\epsilon $ is covariant, and if $i^k$ is covariant, than the phase space distribution
$f$ must also be covariant. Rewriting Equation (3) in the form
\begin{equation}
 \epsilon \epsilon^\prime w f f^\prime (d^3 p/\epsilon) (d^3 p^\prime / \epsilon ^\prime)\; d^3 x\;,
\end{equation}
we reason that, since the phase-space density, $d^3 p / \epsilon$, and $i^k$ are covariant, 
$ \epsilon \epsilon^\prime w$ must also be invariant. It follows that the integral 
\begin{equation}
 W^{k l} = \frac{1}{2} \epsilon \epsilon^\prime \int q^k q^l v_{rel}\; d\sigma\;,
\end{equation}
corresponding to Equation (12), must form a symmetric 4-tensor. Here $v_{rel}$ is the velocity 
of one particle in the rest frame of the other.
Following a derivation similar to (10-13), we find
 \begin{equation}
 B_{\alpha\beta} =  W^{\alpha\beta} / \epsilon \epsilon^\prime = 
  	A \frac{m_e c^2}{\epsilon}
        \frac { \bigg( v^{\prime 2} \delta_{\alpha \beta} - v^\prime_\alpha v^\prime _\beta \bigg)} {v^{\prime 3} }
\end{equation}
to be the desired manifestly covariant generalization of normal kinetic theory.

It is reasonable to neglect the zeroth component of (35). The reason for this is that the change in 
particle energy $q^0$ is of second order with respect to small scattering angle, and thus
$W^{0 0}$ and $W^{0 \alpha}$ are of third or fourth order.  Yet the Fokker-Planck equation is 
accurate only as far as second-order quantities.  Leaving only the components of Equation (8)
corresponding to momentum flux, and re-expressing $B_{\alpha \beta}$ in a convenient form,
we arrive at
\\
\begin{equation}
 C = \frac{\partial}{\partial u_\alpha}\bigg[D_{\alpha\beta}\frac{\partial f(\mathbf{u})}{\partial 
u_\beta}-F_\alpha f(\mathbf{u})\bigg]\;,
\end{equation}
where 
\begin{equation}
 \mathbf{D}_{\alpha\beta} = A
		\int \mathbf{Z}(\mathbf{u},\mathbf{u^\prime}) f(\mathbf{u^\prime})\; d^3 u^\prime\;,
\end{equation}
\begin{equation}
 \mathbf{F} = - A
		\sum_\beta\int\bigg[ \frac{\partial}{\partial u^\prime_\beta}\mathbf{Z}( \mathbf{u},
\mathbf{u^\prime} ) \bigg] f(\mathbf{u^\prime})\; d^3 u^\prime\;,
\end{equation}
and where the kernel is now given by
\begin{equation}
 \mathbf{Z}(\mathbf{u},\mathbf{u^\prime}) = \frac{r^2}{\gamma \gamma^\prime w^3 }
	\bigg[ w^2 \delta_{\alpha\beta} - u_\alpha u_\beta - u_\alpha ^\prime u_\beta ^\prime
	+r( u_\alpha u_\beta ^\prime + u_\alpha ^\prime u_\beta )\bigg]\;.
\end{equation}
Here, $\mathbf{u}$ is the ratio of the momentum to the rest mass $m_0$. 
Also, 
 $\gamma = \sqrt{1+\mathbf{u}^2 / c^2}$, $\gamma^\prime = \sqrt{1+{\mathbf{u}^{\prime}}^2/ c^2}$, 
\begin{equation}
 r = \gamma \gamma^\prime - \mathbf{u} \centerdot \mathbf{u^\prime} / c^2\;,
\end{equation}
and
\begin{equation}
 w = c \sqrt{ r^2 - 1}\;.
\end{equation}
In the non-relativistic limit, $r \rightarrow 1$, $w \rightarrow |\mathbf{u}-\mathbf{u^\prime}|$
and $\mathbf{u} \rightarrow \mathbf{v}$. Thus relativistic kinetic theory, which begins as manifestly
covariant, also transitions to the non-relativistic limit at low velocity. From now on, when we refer to 
`velocity' in a relativistic context, we mean $u = p/m_0$. 

Equation (36) is valid as long as bremsstrahlung is not so dominant as to make particle collisions inelastic,
thus affecting the cross section. Provided this is the case, it is simple enough to calculate the bremsstrahlung 
losses as an integral over the electron distribution.
 
Equations (37) and (38) may be solved with direct numerical integration. The Langevin Equation (28) can then
be used to update a large number of phase-space points, rather than obtaining an implicit solution of the 
differential Equation (6). Besides the benefits of speed and the guaranteed conservation of particle number, 
this approach also allows various new acceleration mechanisms to be incorporated: 
as an example we include diffusion due to Alfv\'enic turbulence and radiative losses within a single-species electron plasma
in \S\S 3 and 4.  It remains to show that the approach conserves energy and that the
energy equipartitions correctly.

\section{Comparison with Existing Theory}

Let us compare this result to that of Nayakshin and Melia (1998, NM98). Since NM98 construct
a kinetic theory around the unitless energy $E = \gamma - 1$, while covariant theory measures the
friction and diffusion coefficients in the generalized velocity $\mathbf{u}=\gamma \mathbf{v}$, 
direct comparison is not
obvious. For example, in NM98 the energy exchange and viscous coefficients are equivalent, whereas 
covariant theory finds the energy exchange by integrating the flux of energy through a volume of phase space.
Covariant theory gives diffusion coefficients both parallel and perpendicular to a particle's velocity;
the diffusion coefficients of NM98, measured in the scalar $\gamma$, do not.

While the two theories measure different quantities, we may nevertheless place them in the same system of units
for direct comparison. The unitless diffusion coefficient is
\begin{equation}
 {D}_{\alpha \alpha} = \frac{A}{c^2}
                \int {Z}(u,u^\prime) f({u^\prime})\; 4 \pi u^{\prime 2} d u^\prime\;,
\end{equation}
where the $c^2$ term gives the diffusion coefficient in units of $1/s$, as in NM98 (the kernel Z is measured
as $1/c$, the constant A as $c^3/s$). With a change of the variable of integration,
\begin{equation}
 {D}_{\alpha \alpha} = \frac{4 \pi A}{c^4}
                \int Z(u,u^\prime) f(\gamma^\prime)\;
                \frac{u^{\prime 3}}{\gamma^\prime} d \gamma^\prime,
\end{equation}
the function
\begin{equation}
D(u, u^\prime) = \frac{4 \pi A}{c^4} Z(u, u^\prime) \frac{u^{\prime 3}}{\gamma^\prime}
\end{equation}
is now the equivalent of Equation (35) in NM98. As we shall see below, the difference
between these two coefficients appears to be the reason why we do not recover Blasi's
(2000) results, since the method of NM98 is only valid in the highly relativistic regime (see
below).

At first, the most glaring discrepancy is the absolute scale of the two theories: covariant theory 
(along with nonrelativistic Rosenbluth potentials) is divergent (Fig. 1a),
while that of NM98 (Fig. 1b) is not. 

There are also physically grounded discrepancies.
We show in the appendix that the Fokker-Planck and dynamical friction approaches are equivalent.
In the latter theory, it is physically clear that all particles must be slowed by the viscous term,
since a moving particle is always struck more frequently on its
leading side, regardless of what relation its speed has to the rest of the group.
To use an analogy with another $1/r$ potential: a star is never accelerated by gravitational drag.
Yet the energy exchange term in NM98 (Fig. 2) changes sign, drawing all particles to the average (scalar) speed.
In this case, one must multiply the frequency of collisions by the fractional change in energy,
which allows particles to both gain and lose energy as the result of multiple collisions. Thus, the covariant
friction is positive-definite, while the `friction' term $a(\gamma, \gamma^\prime)$ of NM98 is not.

Perhaps the best way to compare the two theories, and to gauge their domain of validity, is 
to observe their predictions for the temporal evolution of a distribution. Crucially to the 
application of Section 4, we find that the theory of NM98 may not be applied in the extreme 
non-relativistic limit.

We follow the description of NM98, beginning with a Maxwellian distribution and solving the temporal 
evolution of the distribution via the Chang-Cooper implicit method recommended by Petrosian and Park 
(1996), for a variety of transrelativistic temperatures $\theta\equiv kT/m_ec^2$ (so that $\theta 
\sim 10$ is roughly $5\times10^{10}$ K).  A fully relativistic distribution ($\theta \sim 5$) remains 
in equilibrium, conserving energy and particle number to around $1\%$. The kinetics of NM98 are 
therefore robust for highly energetic particles.

As the temperature falls near the electron rest mass energy, however,
the distribution deviates by between $3$ and $15 \%$. At these intermediate temperatures $\theta \sim 1$,
an algorithm which explicitly conserves particle number (as suggested in NM98 Appendix A) is required 
to resolve the coefficients.  Neither the stochastic nor the implicit algorithms maintain a 
transrelativistic distribution in equilibrium.

At $\theta \sim 0.1$, the distribution ceases to be relativistic, the function exp$(-\gamma/\theta)$
suffers from underflow errors, and the kinetics of NM98 do not maintain an equilibrium distribution.
In Blasi (2000), the distribution 
prior to heating lies at $\theta = 0.01$; it is therefore the failure of NM98 to correctly 
transition to the nonrelativistic limit that leads to the inaccuracy in Blasi's (2000) formulation 
of the cluster's nonthermal emissivity.

\section{A Solution of the Equations in Covariant Kinetic Theory}
Figure 3 shows the kinetic coefficients calculated for a Maxwellian distribution
\begin{equation}
f(v) = \bigg( \frac{m}{2 \pi k T} \bigg)^{3/2}\exp(-mv^2/2kT)\;,
\end{equation}
at a temperature $10^8$ K, compared to the known analytic versions given by Equations (21-23). 
It's important to stress here that, while most
of our results are presented as functions of the magnitude of the velocity, this quantity plays no role in the calculation itself.

To make this figure, a large number (here, one billion) of particles are first initialized, scaled,
and interpolated onto a grid, generally using the cloud-in-cell or triangle-shaped-cloud interpolants well-known from
many-body mesh calculations. This three dimensional grid represents a discrete version of the distribution $f(\mathbf{v}^\prime)$
on the right hand side of Equations (15) and (16). For any velocity $\mathbf{v^\prime}$, a subfunction 
interpolates values off the grid, and the result is an apparently continuous function. 

We now concern ourselves with a single value of $\mathbf{v}$---the left hand side of Equation (15)---and carry out a Monte Carlo
integration of the kernel (Eq. 10) using the Vegas algorithm of the Cuba integration library. 
A maximum of 2500 function evaluations,
beginning with 300 subdivisions and adding 500 new subdivisions in a refinement step gives the best speed-to-accuracy ratio.
For this one particular x, y, and z velocity, we sum over $\beta$, resulting in six independent diffusion coefficients and three
force coefficients comprising a structure.

We do this for some number of $\mathbf{v}$ values, creating a second grid, each point of which is a structure containing
the six $D_{\alpha \beta}$ or three $F_\alpha$. Two computational points are relevant here: first, 
we perform the sum over $\beta$ within the function call, rather than integrating three times and adding the answers. 
Second, the algorithm is about ten times 
faster when a vector of many functions is passed to the integrator
in a single integration call, rather than defining the function and calling the integrator many separate times. 
A lookup table is created so that in calling the integrator to function number one, say, 
the code automatically knows that it's looking for the $D_{xx}$ value for $\mathbf{v} = (1\times
10^8, -2\times 10^8, 2\times 10^8)$ m s$^{-1}$.

Again, so long as the grid is sufficiently fine, a subfunction interpolates $D_{\alpha \beta}$ and $F_\alpha$
off the grid for any $\mathbf{v}$. We now have an apparently continuous function solving (Eqs. 19-20).

The $v$-grid points are divided evenly among several processors using MPI. Each process accumulates its local distribution,
which is then summed and re-broadcast to all processes. A process solves the $\mathbf{v}$ values belonging to it, and finally 
the full grid is gathered. In this fashion, one can avoid the passing of particles.

Finally, for display purposes, we find the magnitude of each particle's velocity and friction, and we diagonalize its
diffusion tensor. That is, the coefficients in Figure 1 play no role in evolving the distribution; the solution
makes no isotropic assumption and is naturally calculated in the lab frame.

To evolve the distribution, each of the particles is updated using a first-order stochastic equation
solver. 
Second order schemes exist (Qiang 2000), but the multiple function evaluations
are prohibitively expensive. 
The tensor Q is found by diagonalizing D via Jacobi rotation, taking the square root of the diagonal components,
and then transforming back.  

Making a transition to a relativistic Maxwellian plasma, we instead use the distribution function 
\begin{equation}
f(u) = \frac{m n_e}{4 \pi k T c\; K_2(m_e c^2/kT)} \exp(- \gamma m_e c^2/kT)\;,
\end{equation}
where $K_2$ is the modified Hankel function
of order 2. Again, we are working with $u=p/m_0$, the relativistic
generalization of the velocity, and we use the term `velocity' exclusively with this meaning. 
With a substitution of $f(u)$ in Equation (37),
the resulting relativistic kinetic coefficients may be defined as
\begin{equation}
D_{R}(u) = \phi(u) D_{NR}(u)\;,
\end{equation}
\begin{equation}
F_R(u) = \psi(u) F_{NR}(u).
\end{equation}
For ease of use, we have expanded $\phi(u)$ and $\psi(u)$---functions we call the {\it enhancement} factors,
these being the multiplicative differences between the non-relativistic and relativistic expressions---as 
\begin{equation}
\phi(u) = \sum_0^3 a_i u^i\;,
\end{equation}
and similarly for $\psi(u)$,
where again the coefficients $a_i$ are expanded as functions of temperature,
\begin{equation}
a_i = \sum_0^8 b_j \bigg( \frac{T}{5\times 10^8\;\hbox{K}} \bigg)^j.
\end{equation}
The resulting coefficients are given in Table 1; the polynomial functions over $b_j$ fitted to each $a_i$ are
shown in Figure 4. In Figure 5 we show the actual enhancement factors $\phi(u)$ and $\psi(u)$, 
together with the polynomials reconstructed from Table 1. Thus, with a few numbers, we can readily 
calculate the relativistic kinetic coefficients.

These enhancements can now be used to redefine the equilibration and viscous time scales, one of our
primary goals in this transition to a covariant kinetic theory. We have
\begin{equation}
D_{\bot \,R} \, t_{d, R} = u^2\;,
\end{equation}
where the average velocity is evaluated from $(\gamma - 1) m_ec^2 = (3/2) \alpha kT$.
The factor $\alpha$ is $1$ in the non-relativistic regime, but approaches $2$ when $\gamma>>1$. 
We may estimate it as the piecewise function
\begin{equation}
\alpha = 
\begin{cases}
	1 & \text{log}_{10}(T)<8.5,\\
	-2.8+0.45 \times \text{log}_{10}(T) & 8.5<\text{log}_{10}(T)<10.5\\
	2 & \text{log}_{10}(T)>10.5\;\;\;\;.
\end{cases}
\end{equation}
Now the length of time required for a relativistic distribution to relax to equilibrium
may be measured against that for a non-relativistic distribution by the ratio 
$t_{d,R}/t_{d, NR} = v^2/u^2 \phi(u)$, which is plotted in Figure 6. (Again, the definition
of the Spitzer time is valid at the average scalar velocity of the group. The Spitzer time
is not particularly useful for calculations: rather, it is designed to
give an intuitive grasp of the amount of time needed for equilibration, and we give its relativistic
generalization as motivation for the re-examination of a wide class of problems.)
Note that a fully relativistic plasma equilibrates in 1/100th of a Spitzer time, $t_s$.

The evolution of the non-relativistic and relativistic distributions---at $10^8$ and $2.5\times 10^9$ 
K, respectively---is shown in Figures 8 and 9, as a function of the Spitzer time, $t_s$. The 
non-relativistic distribution requires about two Spitzer times to equilibrate, whereas 
the relativistic one reaches equilibrium in only $0.2\,t_s$.

When the energy is constant, or when the injection rate is known, we use normalization 
coefficients to maintain constant energy in the distribution; an example of how these
coefficients vary with time is plotted in Figure 10. Energy is conserved to an accuracy of
$10^{-2}$, about the same as the precision of our integration. 

However, when energy is not constant, this simple procedure is not feasible, and we must 
increase the grid size to a minimum of $64^3$ and the maximum number of function evaluations
to 10,000. In this case, normalization is no longer required. For example, in Figure 11, we allowed 
the distribution to relax for a full Spitzer time before turning on stochastic acceleration,
conserving energy all the while.

\section{Sample Application to Cosmic Ray Acceleration in Galaxy Clusters}

Evidence for the presence of relativistic electrons in the intracluster medium (ICM) is provided primarily
by diffuse synchrotron emission at radio wavelengths.
Given the $\sim$ Mpc-size of the synchrotron
features, and the fact that the radiative lifetime of the electrons appears to be orders of magnitude
shorter than the time required to cover such distances,
the presence of this emission also suggests that electrons are being accelerated out of
the ICM thermal pool. 

Typically, it is assumed that the radiating
particles must be continuously re-accelerated on their way out---these are `primary' models 
(Jaffe 1977; Schlickeiser et al. 1987, Brunetti et al. 2001, Ohno et al. 2002).
However, protons could diffuse throughout the cluster without
radiating, all the while colliding with thermal ICM ions and cascading into the observed relativistic electrons---these
are `secondary' models (Dennison 1980). Secondary models relieve the difficulties of continuously accelerating
electrons against radiative and Coulomb cooling, however Blasi and Colafrancesco (1999) suggest that protons could
not provide enough secondary electrons to describe the radio emission without also producing $\pi^0$ gamma
decays in excess of the number observed by EGRET. 

Coma's radio emission is accompanied by hard X-rays of similar spatial distribution; this and
its regular morphology classify the Coma as a `radio halo'. In contrast, `radio relics' are typically irregular and
concentrated toward the cluster's periphery (e.g., Feretti 2003). 

Interpreted as thermal bremsstrahlung,
the observed X-rays imply an ICM temperature between 8 and 9 keV. Coma's X-ray emission cannot be described entirely 
as thermal bremsstrahlung, however.
The Rossi X-Ray Timing Explorer (Rephaeli and Gruber 2002)
and BeppoSAX (Fusco-Femiano et al. 2004, hereafter FF04) have each made two observations of the Coma, both claiming
to find a hard X-ray tail (HXR) in excess of a thermal flux. This is a controversial claim:
the initial analysis of Fusco-Femiano et al. (1999, FF99)
was flawed---one of three spectra was counted twice---leading to a ~2.5 $\sigma$
overstatement of the confidence level, according to Rossetti and Molendi (2004, RM04). RM04 flatly state that the
second BeppoSAX observation shows no evidence of a hard tail. FF04, however, find a significant drop in the 
$\chi ^2$ value for the fit to a hybrid thermal-power-law model ($\chi^2 = 1.20$
for 7 dof) as opposed to the purely thermal model ($\chi^2 = 4.10$ for 9 dof). This improvement cannot be
matched by a two-temperature model, since the second temperature ($kT>50$ keV) is considered unrealistic.

Rephaeli (1979) had predicted
such a tail must exist when the synchrotron-emitting electrons inverse Compton scatter with the cosmic background. 
In principle, this origin could be used as a probe for the ICM magnetic field.
Yet, assuming that both the synchrotron radio and Compton HXR are produced by the same population of electrons,
FF04 infer a magnetic field of $B \sim 0.2\; \mu$G---more than a factor of ten below the results inferred by Faraday
rotation ($B \sim 6.0\;\mu$G). 

For this reason, nonthermal bremsstrahlung
(Blasi 2000) was proposed as an alternative emission mechanism to inverse Compton. Here, nonthermal electrons
are directly accelerated and directly radiate via $e-e$ bremsstrahlung.
Petrosian (2001) has argued that bremsstrahlung is not sufficiently efficient to cool the distribution
before an unacceptably large amount of energy is given to the ICM, shifting the thermal body of bremsstrahlung
above its observed flux. Because this problem is simple (it requires only stochastic acceleration, 
bremsstrahlung and Coulomb cooling), and because it straddles 
the transrelativistic regime which our theory handles uniquely,
we have chosen it as an example of the technique's power and simplicity.
                                                                                                                                                 
Our calculations agree with the analysis of Petrosian. They suggest that the source of confusion is the inability of
NM98 and DL89's kinetics to reduce to the nonrelativistic limit.
We find that the
stochastically gained energy heats the body of the distribution,
not just the tail, on a Spitzer timescale of some tenths of a Myr; compare this to the tenths of Gyr required for Blasi's model.
Similarly, the Spitzer time requires that any merger event must have occurred within a few Myrs, a
vanishingly small window of time.

With this motivation in mind, let us state the problem more precisely. We use the pitch-angle averaged diffusion 
coefficient (Dermer, Miller, and Li 1996),
\begin{equation}
D(u) = \frac{\pi}{2} \bigg[ \frac{q-1}{q(q+2)} \bigg] c^3 k_0 \beta_A^2 \eta_A (r_B k_0)^{q-2} p^q / \beta \;,
\end{equation}
to represent the resonant interaction of particles with Alfv\'en waves. Here
$p = \gamma \beta$, $r_B = m_ec^2/eB$, $\eta_A = 0.07$ is the fraction of
magnetic energy in Alfv\'en waves, $q=5/3$ is the Kolmogorov constant, 
$L_c = 10$ kpc is the size of the galaxy, $\beta_A = v_A/c$ (where
$v_A = B/ \sqrt{4\pi n_p m_p}$ is the Alfv\'en velocity), and $k_o = 2 \pi / L_c$
is the largest-scale wavenumber. Our point of departure is that of Blasi's calculation
($n = 4\times 10^{-4}$ cm$^{-3}$, $B = 0.8\;\mu$G, and $T = 7.5$ keV), but we also look at a small
range around these figures.

In Figure 10, we show the evolution of the 
distribution, begun at 7.5 keV, with a window of 0.1 Spitzer times between each curve. One full Spitzer 
time elapses before stochastic acceleration from a $0.8\;\mu$G field is turned on (corresponding to the
first curve on the left). We see that rather than stochastic acceleration producing a high-energy tail, 
the high rate of thermalization provides energy to the whole distribution.  

The emissivity of this distribution is calculated self-consistently using particle number distributions
and electron-proton bremsstrahlung cross section,

\begin{equation}
 \frac{dE}{dV\,dt\,d\nu} \equiv j = hkcn_en_z 
	{\int^{ \infty }_{ \sqrt{2k} } \frac{ p_o }{ \sqrt{1+p_o^2} } \frac{d \sigma}{d k} f(p_o) d^3p_o}\;\bigg.\bigg/\;
	{\int^{ \infty }_{ 0 } f(p_o) d^3p_o}\;,
\end{equation}
where ${d \sigma}/{d k}$ is the relativistic bremsstrahlung cross section
averaged over all angles, and expanded to the $(p_o)^6$ th order. Including the
Elwert correction factor, this is (Haug 1997)
\begin{align}
 \frac{d \sigma}{d k} \approx&\frac{ 2 \alpha r_o^2 }{ k p_o^2} \bigg[ \frac{4}{3} \epsilon_o \epsilon_f
			+k^2-\frac{7}{15} \frac{k^2}{ \epsilon_o \epsilon_f } 
			- \frac {11}{70} \frac{ k^2(p_o^2+p_f^2) }
        		{ (\epsilon_o \epsilon_f) ^ 4 }\bigg]\times\nonumber\\
			& \bigg[2ln (\frac{ \epsilon_o \epsilon_f + p_o p_f - 1}{k}) 
			       - \frac{p_o p_f}{ \epsilon_o \epsilon_f}\times\nonumber \\
			& \bigg( 1 + \frac{1}{ \epsilon_o \epsilon_f} + \frac{7}{20} \frac {p_o^2 + p_f^2}
			{ (\epsilon_o \epsilon_f)^3 } + (\frac{9}{28} k^2 + \frac{263}{210} p_o^2p_f^2)
			\frac{1}{ (\epsilon_o \epsilon_f)^3 } \bigg) \bigg]\times\nonumber \\
			& \frac{a_f}{a_o} \frac{ 1-exp(-2 \pi a_o) }{ 1-exp(-2 \pi a_f) }\;\;,
\end{align}
where $ \alpha = e^2/ \hbar c$ is the fine structure constant, $r_o = e^2/m_e c^2$ is the electron radius,
$k = \hbar \omega / m_e c^2$ is the photon energy in units of the electron mass energy, $p_o = \gamma_o \beta_o = u_o/ c$
is the initial electron momentum in units of $m_e c$, $ \epsilon_o = \gamma_o $ is the total energy in rest mass units, $a_o = \alpha \epsilon_o / p_o$, and
\begin{equation}
p_f = \bigg( k^2 - p_o - 2k\sqrt{1-p_o^2} \bigg)^{1/2}\;.
\end{equation}
In the energy regime considered, the error of this expansion is a few percent---the contribution of $e^-$--$e^-$
and $e^-$--$e^+$ bremsstrahlung is of this order, so we exclude them from consideration.

Once the cross section in Equation (55) is found, we calculate the observed X-ray flux $F_x$ under 
the assumption of constant density and use the accepted volume $V$ and distance $d_L$ for Coma to
write $F_x = V j/(4 \pi d_L^2)$, assuming a (dimensionless) Hubble constant $h=0.6$.
The calculated spectrum is shown in Figure 11, together with the \emph{BeppoSAX} data, at spacings of 
$0.8 t_s$. 

For completeness, we also show the evolution for nine combinations of electron density
and magnetic field. In the bottom left corner---higher density and lower magnetic field---a nonequilibrium
tail never heats. In the top right---lower density and higher magnetic field---the soft X-ray flux 
immediately moves above the observed limits. 
In none of these figures do we address the other free parameter, the initial temperature. And,
while the calculation is done via the relativistically correct kernel, in no case is a
relativistic tail produced before the nonrelativistic body heats to some excessively high energy. 

All of these figures evolve over $4 t_s = 2$ Myr, some factor $>10^2$ below that reported by Blasi.
The corrected Spitzer timescale is irreconcilable with the HXR observations (Fig. 11), as in no case
is a nonthermal tail accelerated before the ICM equilibrates the added energy and reaches an 
unacceptably high temperature. This is consistent with the result reported by Petrosian (2001): the 
erstwhile lack of a kinetic theory that is applicable throughout the transrelativistic regime was 
the source of discrepancy.

\section{Concluding Remarks}

We have described a novel method for solving the covariant Landau equation, unconstrained by
assumptions of isotropy or low particle energy. Using simple polynomial fitting formulae for
the kinetic coefficients, we have described an efficient numerical technique for determining
the temporal evolution of an arbitrarily relativistic particle distribution, which may also
be subject to energy injection from the anomalous acceleration of particles, e.g., in shocks or 
scattering with Alfv\'en waves. We anticipate that this method will find widespread application
not only in astrophysics, but other physics disciplines as well.

To demonstrate the power of our technique in understanding the behavior of high-energy plasmas
in astrophysics, we have examined one of the currently considered models---bremsstrahlung emission
by a nonthermal tail---for the hard X-ray emission in galaxy clusters, such as Coma. Our conclusion 
is that the time required by the underlying plasma to attain equilibrium is far too short compared 
to the stochastic acceleration time scale for any nonthermal high-energy extension to survive
longer than $\sim 0.5$ Myr. It seems unlikely, therefore, that cluster mergers could be responsible
for energizing the plasma turbulence required to sustain the nonthermal particle emissivity. 

The application of covariant theory to relativistic electrons in the ICM remains highly relevant.
As we suggest here, the time required for relaxation may vary by orders of magnitude from the 
currently used quantity.  In a future paper, we will examine the production of $\pi^0$ gamma 
rays and the associated cascade of charged leptons during proton-proton collisions in the
ICM. In addition, we have not yet completely resolved the complete set of differences between 
covariant theory and the kinetics of DL89 and NM98, which amounts to better defining the range
of validity for the latter approximate treatments. We are performing a detailed comparison of 
their energy exchange and temporal evolution properties, and these results too will appear in 
a future paper.

\section{Acknowledgments}

This research was supported by NASA grant NAG5-9205 and NSF grant AST-0402502 at
the University of Arizona.

\clearpage\newpage
\section{Appendix A: Equivalence of the Langevin and Fokker Planck approaches}
We argue following Zwanzig (2001). Consider the stochastic equation
\begin{equation}
\frac{d v}{dt} = F_d (v) + \Lambda (t)\;,
\end{equation}
where the noise $\Lambda(t)$ is Gaussian, with zero mean and delta-correlated second moment,
\begin{equation}
 \langle \Lambda(t) \Lambda(t) \rangle = 2B \delta (t - t^\prime)\;.
\end{equation}
Let us now find the probability distribution $f(v,t)$, averaged over the noise. Now, $f(v,t)$ is conserved:
\begin{equation}
 \frac {\partial f}{\partial t} + \frac{\partial}{ \partial v} 
			\bigg( \frac {\partial v}{\partial t} f \bigg) = 0\;. 
\end{equation}
Thus, substituting ${\partial v}/{\partial t}$ we arrive at the stochastic differential equation
\begin{equation}
 \frac {\partial f}{\partial t} = - \frac{\partial}{ \partial v}  
	\bigg[ F_d(v) f(v,t) + \Lambda(t) f(v,t) \bigg] = -Lf - \frac{\partial}{ \partial v} \Lambda(t) f\;,
\end{equation}
where $L\Phi \equiv ({\partial}/{ \partial v}) F_d(v)\Phi$ is an appropriately defined operator.
 The solution to this equation is
\begin{equation}
 f(v,t) = e^{-tL} f(v,0) - \int^t_0 ds\; e^{-(t-s)L} 
	\frac {\partial} {\partial v} \Lambda(s) f(v, s).
\end{equation}
Placing Equation (61) back into Equation (60), we arrive at
\begin{equation}
 \frac {\partial f}{\partial t} = -L f(v,t) - \frac{\partial}{ \partial v} \Lambda(t) f(v,0)
	+ \frac{\partial}{ \partial v} \Lambda(t) \int^t_0 ds e^{-(t-s)L}
	 \frac{\partial}{ \partial v} \Lambda(t) f(v,s)\;,
\end{equation}
which is readily observed to be the Fokker Planck equation, 
\begin{equation}
 \frac {\partial}{\partial t} \langle f(v,t) \rangle = 
	- \frac {\partial}{ \partial v} F_d(v) \langle f(v,t) \rangle
	+ \frac {\partial}{ \partial v} B \frac {\partial}{ \partial v} \langle f(v,t) \rangle
\end{equation}
once the average of $f(v,t)$ is taken, and the zero average and unit correlation of $\Lambda$ is imposed.

\clearpage

\begin{table}[h]
\caption{Expansion coefficients (for Eqs. 46 and 47).\label{tbl-1}}
\footnotesize
\begin{tabular}{crrrrrrrrrrr}
\tableline\tableline\\
$D_\bot$ & $b_0$ & $b_1$ & $b_2$ & $b_3$ & $b_4$ & $b_5$ & $b_6$ & $b_7$ & $b_8$ \\
\tableline
$a_0$ &-1.61e-37 &3.38e-35  &-2.92e-33  &1.34e-31  &-3.50e-30 &5.01e-29  &-2.97e-28 &-1.44e-27 &3.76e-26  \\
$a_1$ &2.72e-28  &-5.76e-26 &5.08e-24   &-2.43e-22 &6.87e-21  &-1.18e-19 &1.20e-18  &-6.91e-18 &9.48e-18   \\
$a_2$ &-7.04e-20 &1.36e-17  &-1.08e-15  &4.48e-14  &-1.07e-12 &1.51e-11  &-1.38e-10 &1.03e-9   &-1.02e-9  \\
$a_3$ &1.74e-11  &-3.37e-9  &2.64e-7    &-1.08e-5  &2.46e-4   &-3.12e-3  &2.10e-2   &-5.77e-2  &1.0716 \\
\tableline\tableline\\
$D_{||}$ & $b_0$ & $b_1$ & $b_2$ & $b_3$ & $b_4$ & $b_5$ & $b_6$ & $b_7$ & $b_8$ \\
\tableline
$a_0$ &4.96e-38  &-1.06e-35 &9.28e-34  &-4.31e-32 &1.13e-30  &-1.61e-29 &9.98e-29  &1.10e-28  &-3.29e-27  \\
$a_1$ &-3.37e-29 &7.51e-27  &-6.91e-25 &3.36e-23  &-9.19e-22 &1.35e-20  &-7.88e-20 &-3.66e-19 &6.35e-18   \\
$a_2$ &8.49e-21  &-2.15e-18 &2.22-16   &-1.21e-14 &3.74e-13  &-6.51e-12 &5.54e-11  &-5.15e-11 &-3.65e-10  \\
$a_3$ &1.55e-12  &-2.90e-10 &2.17e-8   &-8.30e-7  &1.73e-5   &-2.03e-4  &1.60e-3   &2.35e-2   &1.00 \\
\tableline\tableline\\
$F$ & $b_0$ & $b_1$ & $b_2$ & $b_3$ & $b_4$ & $b_5$ & $b_6$ & $b_7$ & $b_8$ \\
\tableline
$a_0$ &1.09e-37  &-2.41e-35 &2.24e-33  &-1.13e-31 &3.39e-30 &-6.17e-29 &6.74e-28 &-4.24e-27 &1.39e-26  \\
$a_1$ &5.80e-29  &-1.24e-26 &1.10e-24  &-5.30e-23 &1.51e-21 &-2.59e-20 &2.66e-19 &-1.62e-18 &7.51e-18   \\
$a_2$ &-7.43e-22 &4.79e-20  &7.39e-18  &-9.70e-16 &4.49e-14 &-1.03e-12 &1.27e-11 &-1.19e-10 &-7.18e-10  \\
$a_3$ &-5.87e-13 &1.04e-10  &-7.20e-9  &2.24e-7   &-1.73e-6 &-7.14e-5  &1.45e-3  &3.72e-2   &9.92e-1 \\
\end{tabular}
\end{table}

\begin{figure*}[ht]
\centering
\plotone{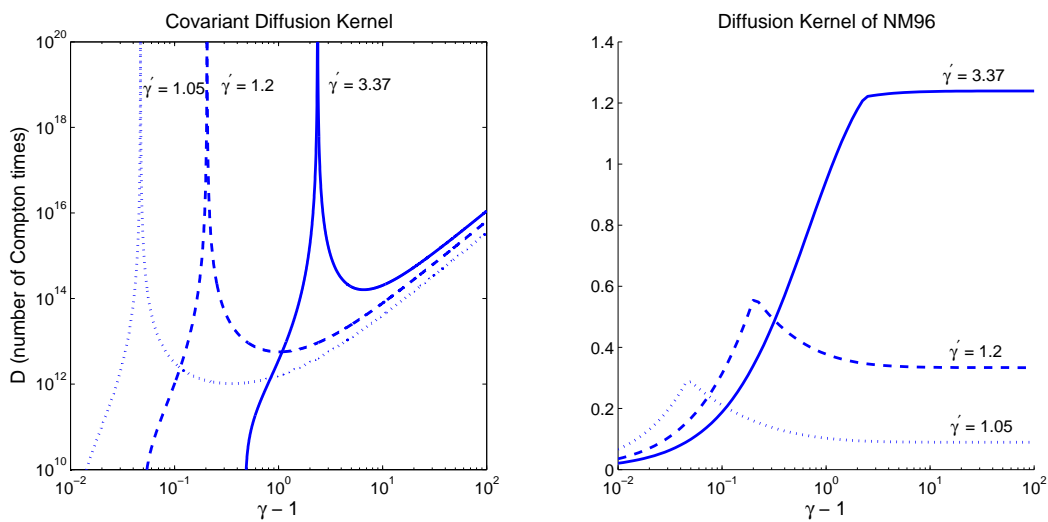}
\caption{A comparison of our theory with that of Nayakshin and Melia 1998. Our kernel (Eq. 44)
is divergent when $\gamma = \gamma^\prime$, while that of NM98 is not. Taking $\gamma^\prime = 1.05$ to be
effectively nonrelativistic, we can see that the theory of NM98 does not duplicate the results of Rosenbluth (1957).
}
\end{figure*}

\begin{figure}
\centering
\plotone{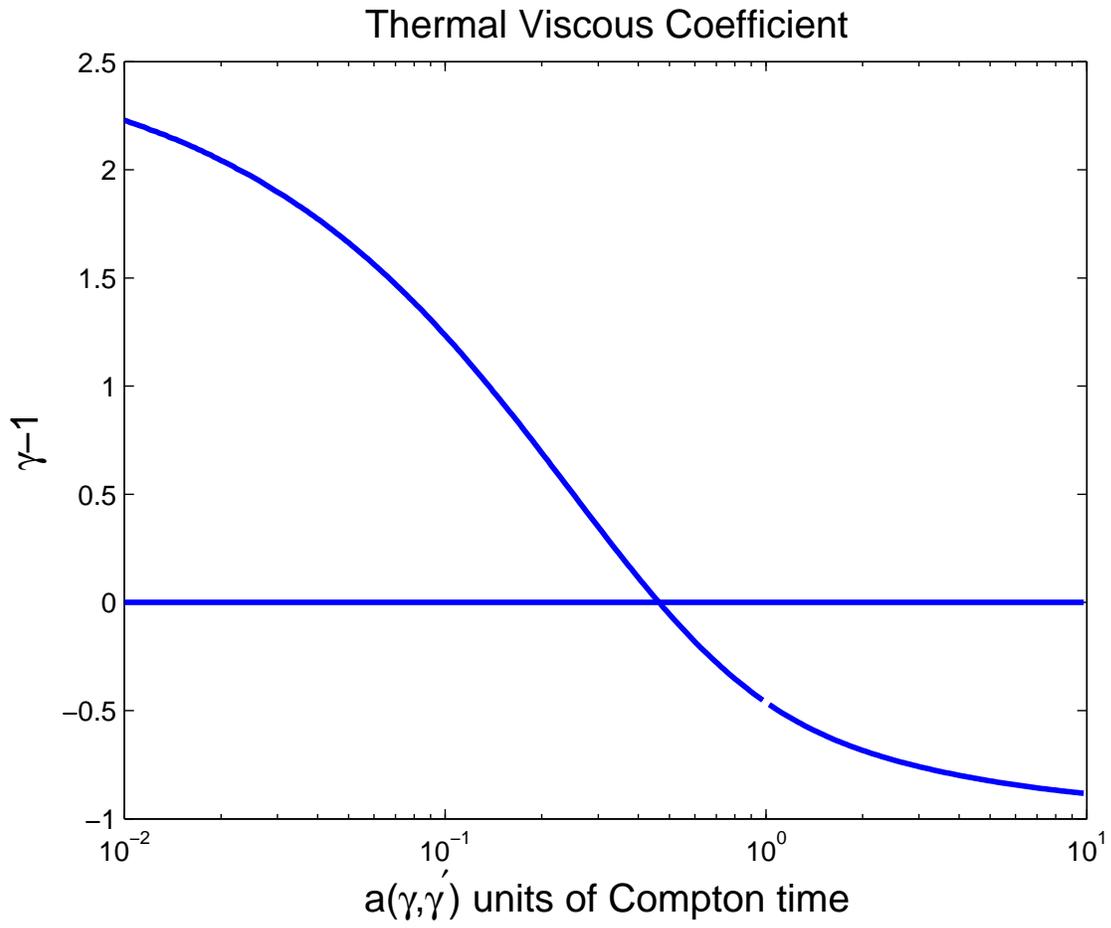}
\caption{
NM98's viscous coefficient $a(\gamma, \gamma^\prime)$ is not positive-definite, as it is in standard kinetic theory (Fig.3).
}
\end{figure}

\begin{figure}
\centering
\plotone{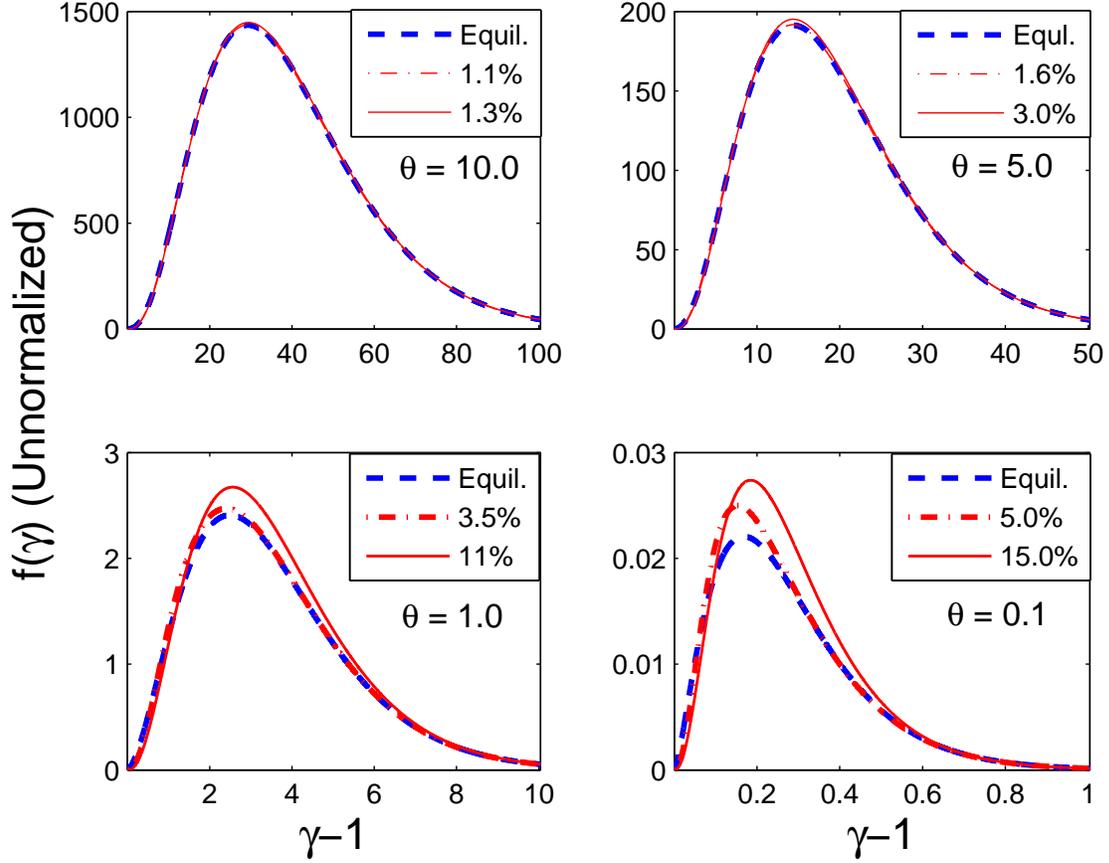}
\caption{
The theory of NM98 breaks down in the non-relativistic limit, perhaps due to limited machine precision as
the Lorentz factor approaches unity. Here, a distribution begun at equilibrium (dashed)
evolves to $1/4$ (dash-dotted) and $4$ (solid) Spitzer times under the influence of NM98's kinetic coefficients.
The quantity $\theta$ is the fraction of thermal energy to electron
rest mass energy; $\theta = 10$ is roughly $5\times10^{10}$ K.
}
\end{figure}

\begin{figure}
\centering
\plotone{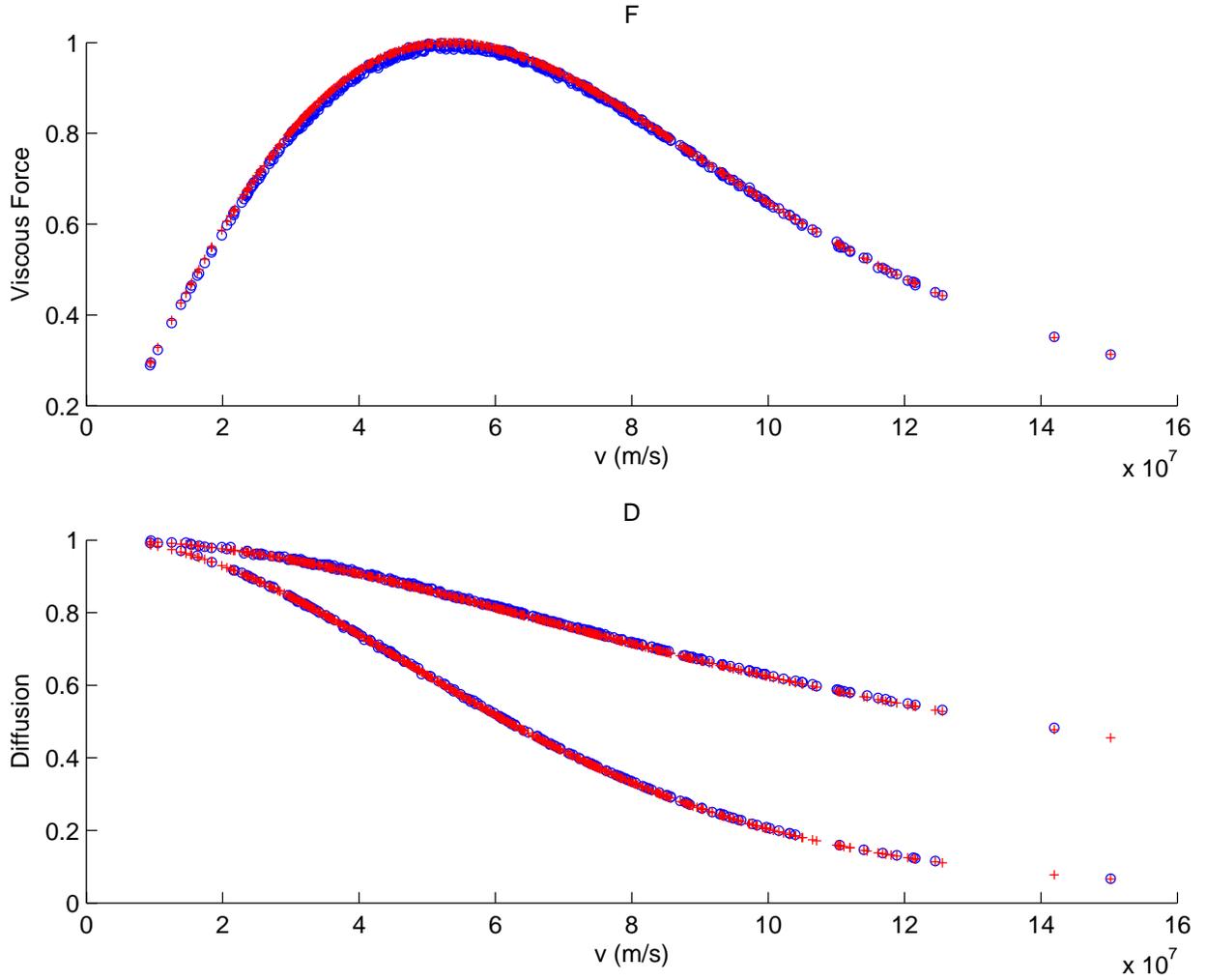}
\caption{Analytic (plus signs) and calculated (circles) $D(v)$ and $F(v)$.
The Viscous Force and the Diffusion coefficients (perpendicular on top, parallel below) 
are normalized to one.}
\end{figure}

\begin{figure}
\centering
\plotone{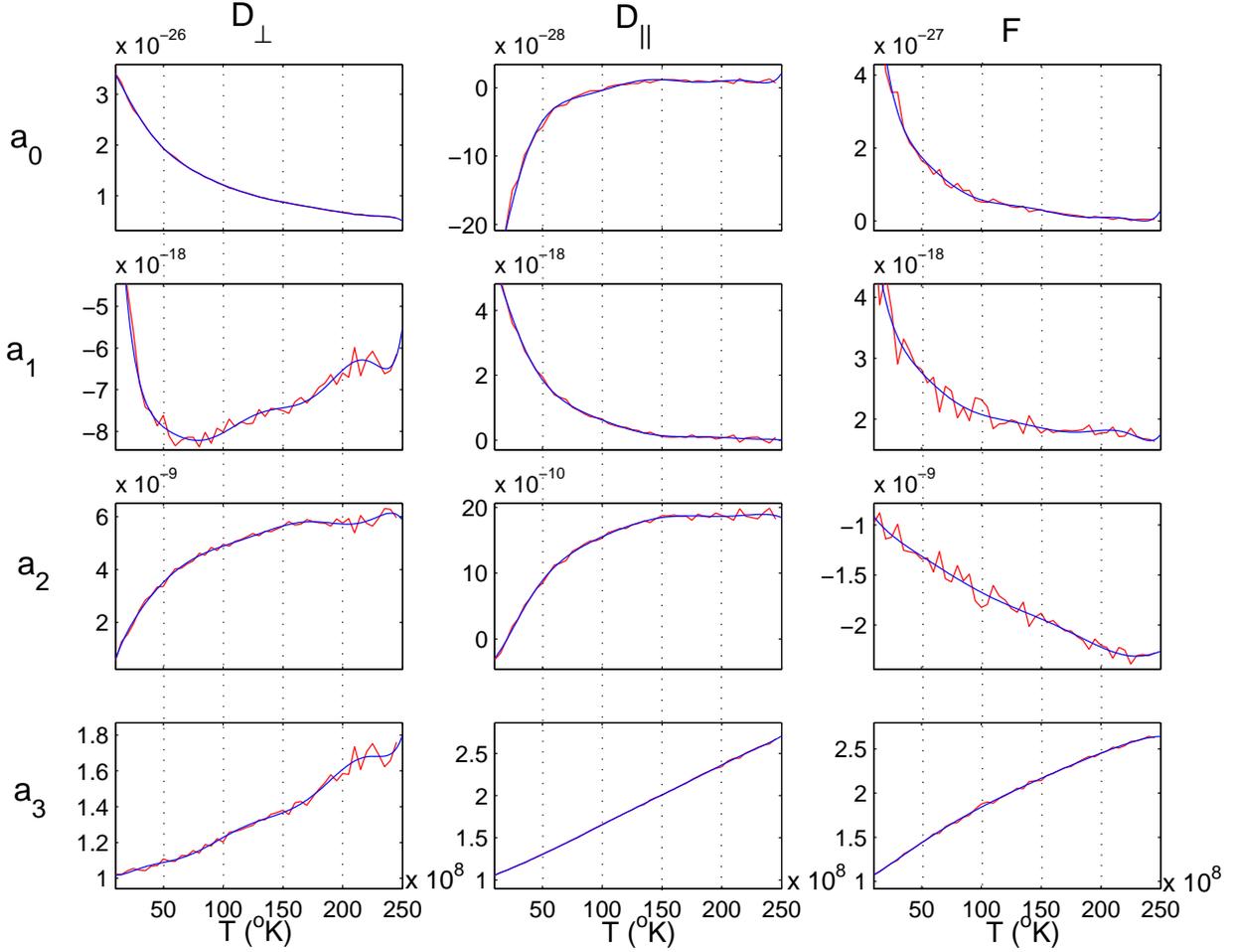}
\caption{The four parameters $a_i$, and the fitting polynomials (shown as smooth continuous
curves) $\sum_0^8 \,b_j\, ( {T}\,/\,{5\times 10^8\;\hbox{K}} )^j$.
We have expanded the kinetic coefficients for a particular temperature---and, 
because this expansion changes as the
temperature is raised, we again expand each of the coefficients as though they were polynomial functions
of T. We present this figure not simply to confirm the accuracy of the expansion, but so that a rough estimate
can be made by reading off the $a_i$ for the temperature of interest.}
\end{figure}

\begin{figure}
\centering
\plotone{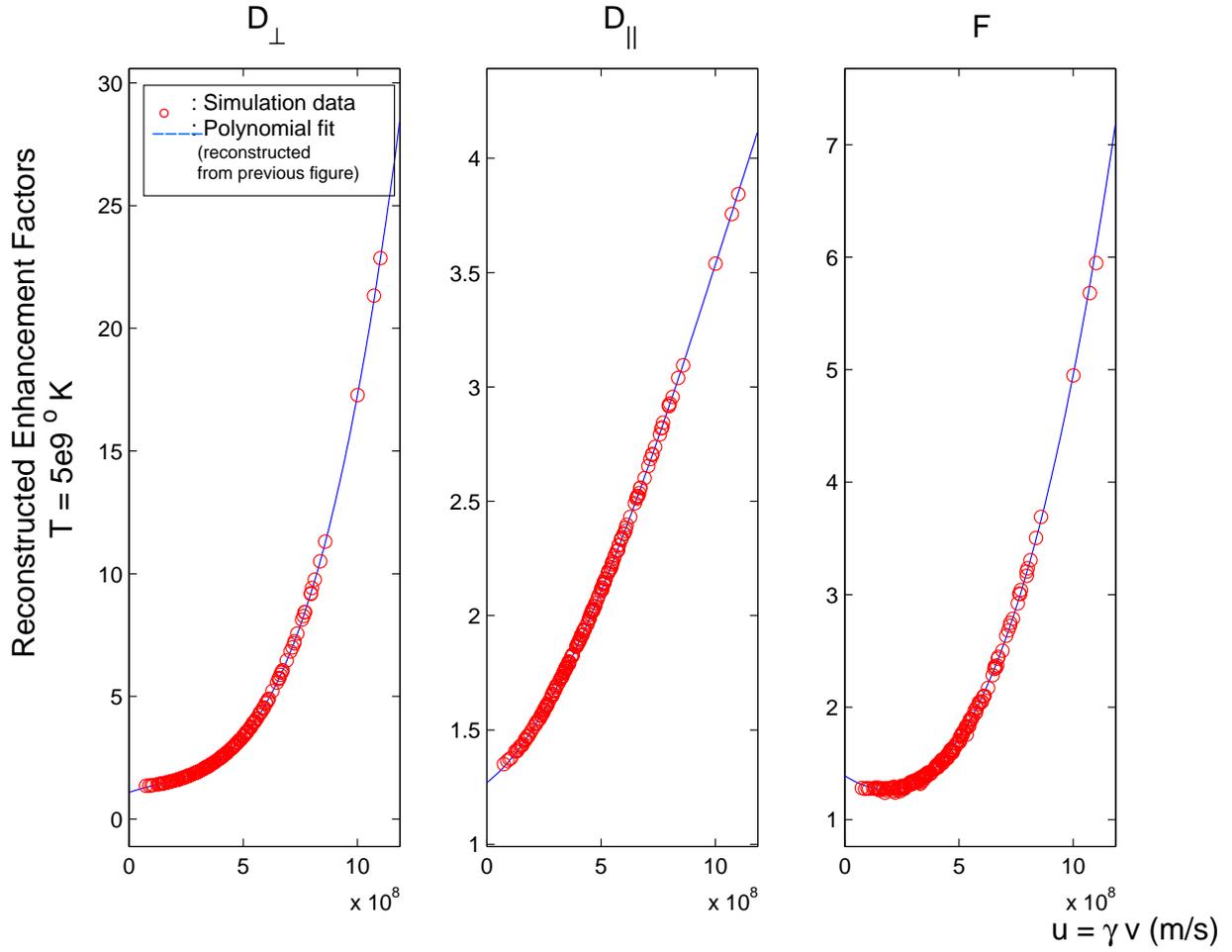}
\caption{Enhancement factors $\phi(u)$ and $\psi(u)$, reconstructed from Table 1, together with the raw data.}
\end{figure}

\begin{figure}
\centering
\plotone{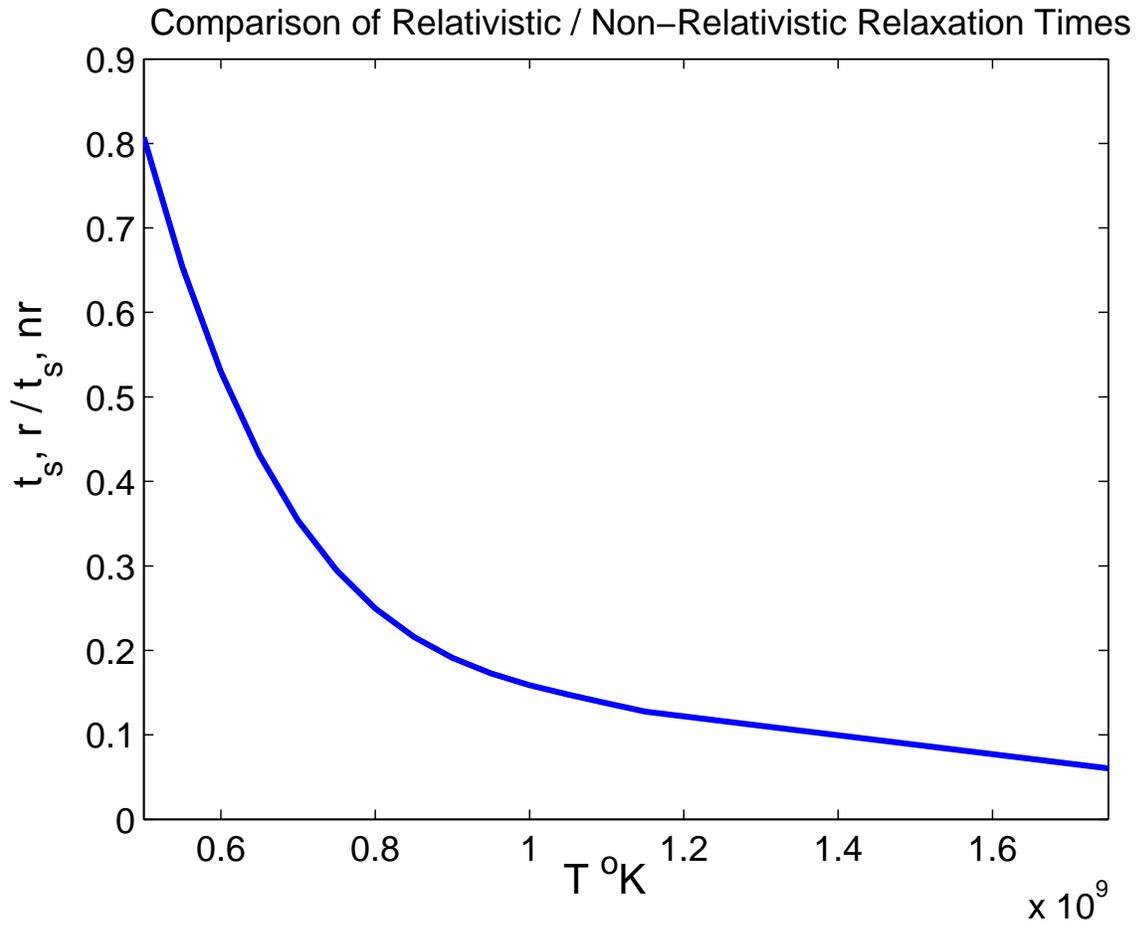}
\caption{Ratio of relativistic to non-relativistic relaxation time scales, as a function of
temperature.}
\end{figure}

\begin{figure}
\centering
\plotone{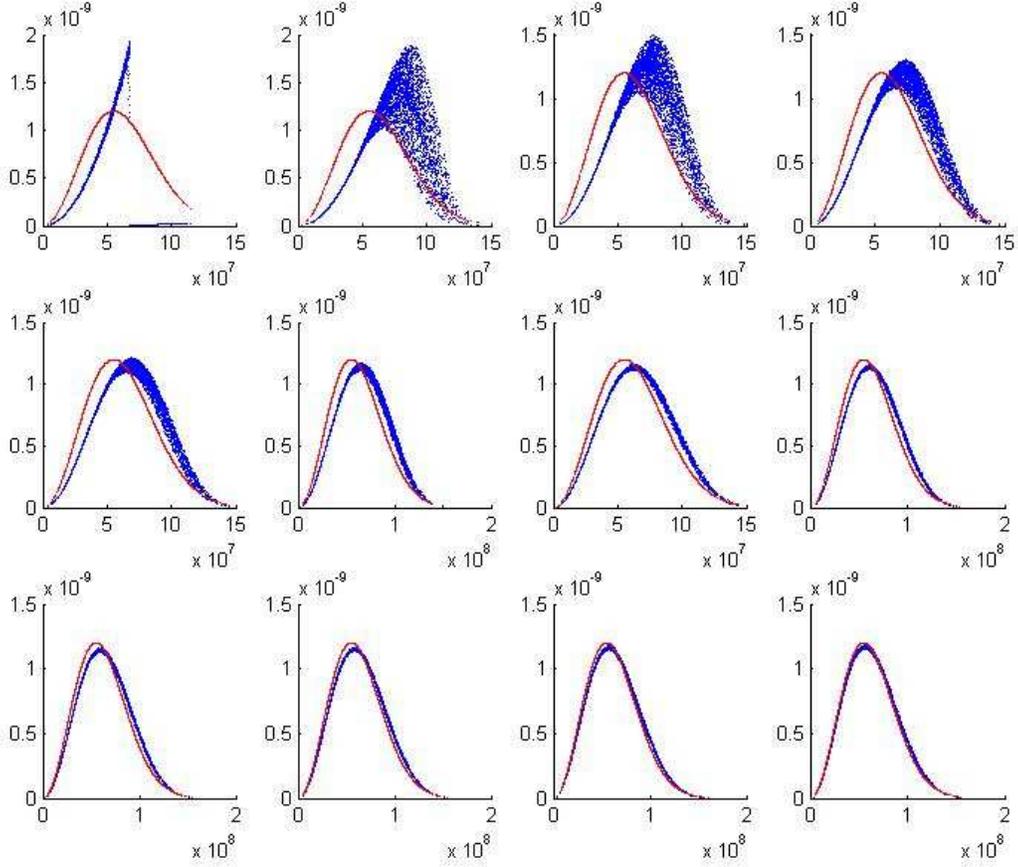}
\caption{A distribution of particles (dotted) initialized with equally likely $v_x, v_y$, and $v_z$,
whose standard deviations are that of a Maxwellian with temperature $T=10^8$ K, relaxes (to the
analytic form, shown as a solid curve) in a time 2.1 $t_s$. The top four frames are in multiples of $0.1t_s$, the
remaining 8 of $0.2t_s$}
\end{figure}

\begin{figure}
\centering
\plotone{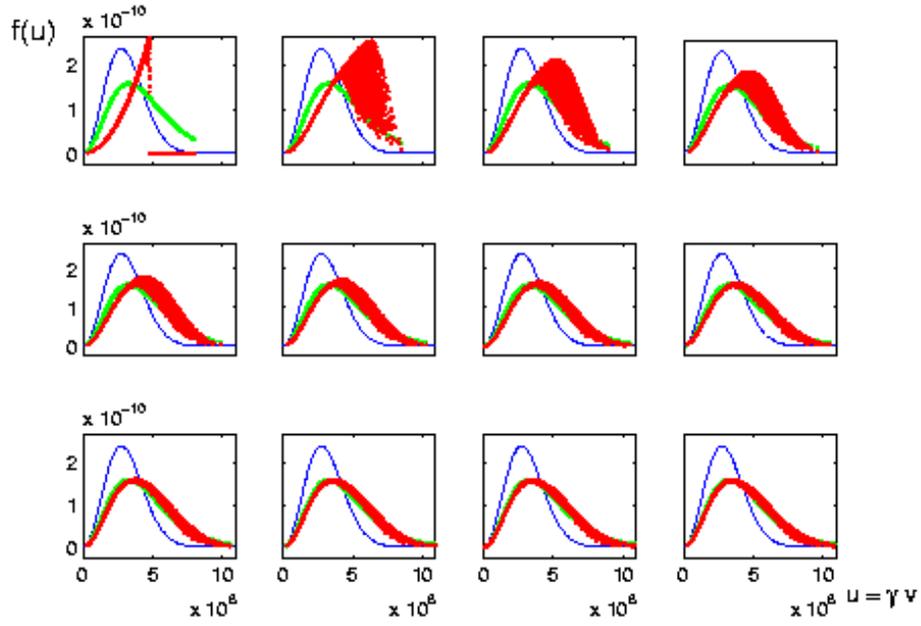}
\caption{An initially cold distribution of particles (dotted) with the same energy as a relativistic 
Maxwellian at $T=2.5\times 10^9$ K (thick solid) relaxes in $0.2\,t_s$. Top four frames are in multiples of $0.015t_s$,
remaining 8 of $0.025t_s$. For comparison, we show the 
non-relativistic distribution (Eq. 42; thin solid) at this temperature.}
\end{figure}

\begin{figure}
\centering
\plotone{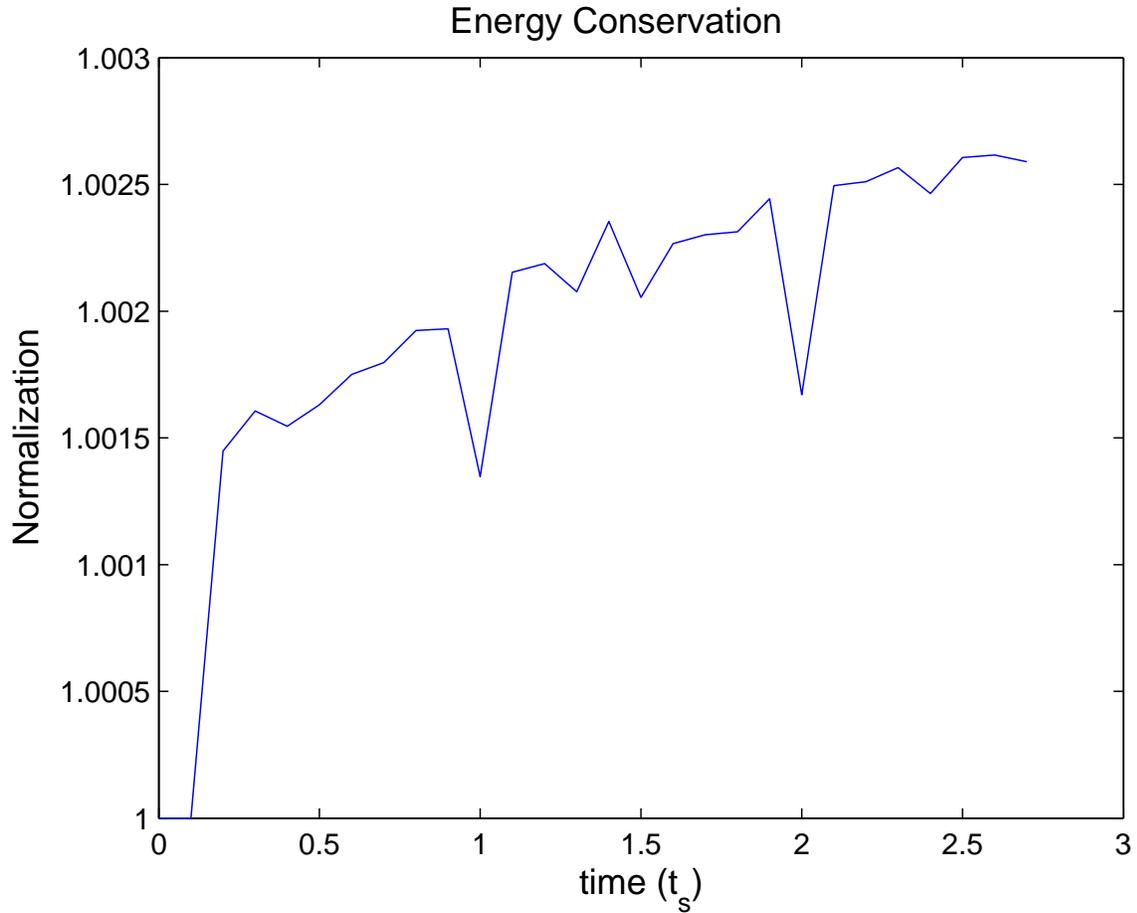}
\caption{For a $32^3$ grid, the kinetic coefficients are correct only
to 2 significant figures; since they are self-consistent 
even this small deviation can lead to run-away heating, and we must ``normalize"
with a slowly evolving coefficient, plotted here as a function of time in units
of the Spitzer time $t_s$. This slight deviation from strict energy conservation can be removed
by using instead a $64^3$ grid (with its accompanying 4 significant figures of accuracy).}
\end{figure}

\begin{figure}
\centering
\plotone{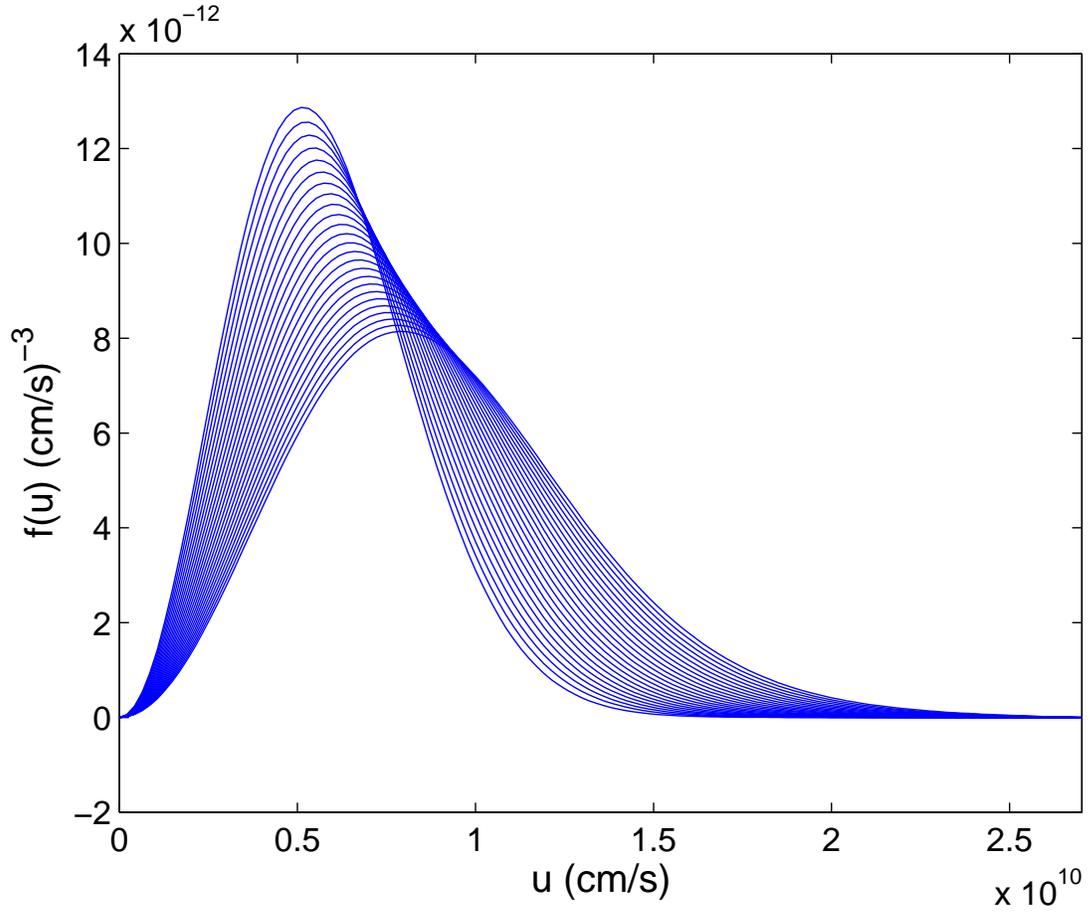}
\caption{The distribution evolves self-consistently from an initial temperature
of $7.5$ keV (curve on the left), as it is heated by Alfv\'enic diffusion and cooled 
by Coulomb diffusion and bremsstrahlung emission (curves moving progressively to the
right) as the entire body heats over a period of only $4t_s = 2$ Myr.}
\end{figure}

\begin{figure}
\centering
\plotone{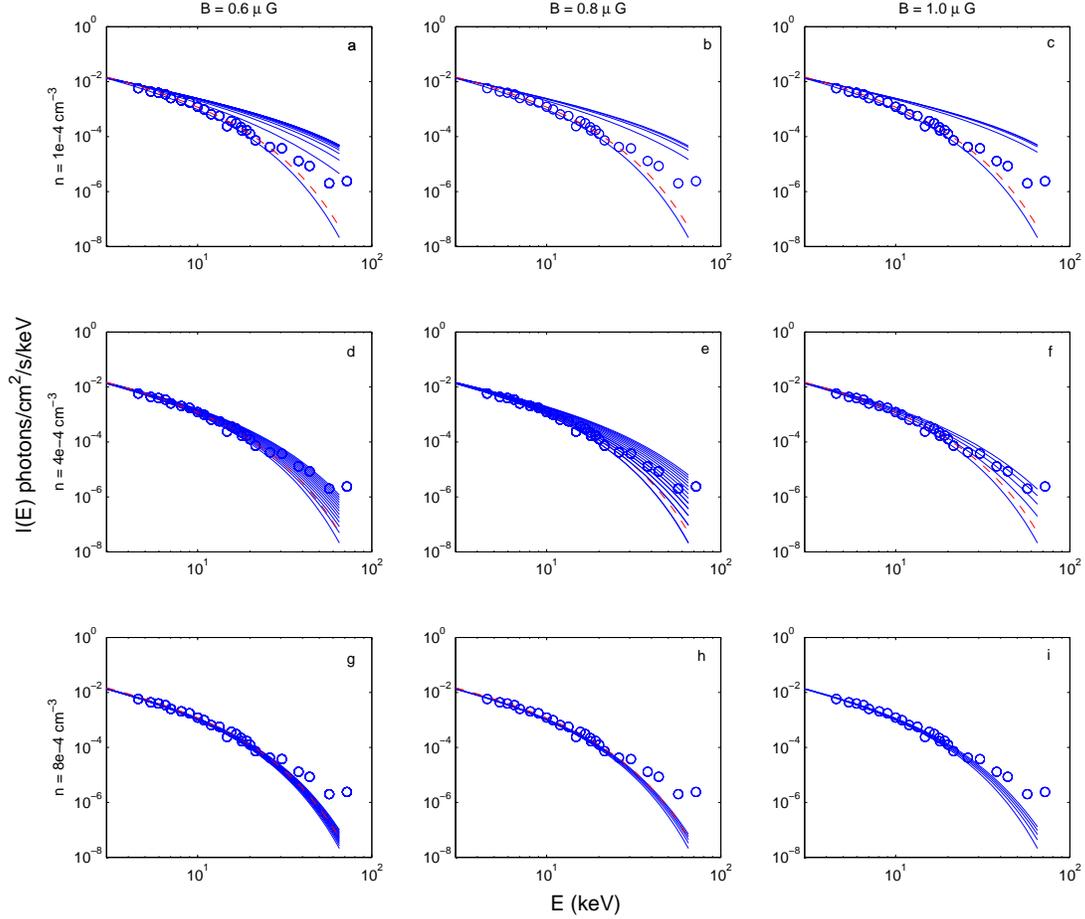}
\caption{X-ray spectra produced by the distributions shown in Figure 10 are in panel e. The others are from a range
of possible values of magnetic field and density. Each line is plotted at an integer of $0.4t_s$. In a,d,g, and e, 
we have allowed $8t_s$ to pass, to show that the body of the distribution heats by the time a nonthermal tail is 
produced; all other figures evolve over $4t_s$.  Data points (error bars not shown) are from the first observation 
of \emph{BeppoSAX} (Fusco-Femiano et al. 1999). Dashed line is equilibrium flux at a temperature of $8.21$ keV. 
Either the distribution is never driven from equilibrium (d,g,h, and i), the body receives an excessive amount of 
heat (e,f) or the whole distribution goes over to a power law (a,b and c).
}
\end{figure}

\end{document}